\documentclass[10pt,aps,prl,twocolumn, superscriptaddress]{revtex4-2}

\usepackage{graphicx}
\usepackage{dcolumn}
\usepackage{bm}
\usepackage{orcidlink}
\begin{document}

\preprint{APS/123-QED}
\title{Anisotropy-induced Inhomogeneous Melting in Finite Dust Clusters}
\author{Sushree Monalisha Sahu\,\orcidlink{0009-0002-7263-702X}}
\email{monalisha.nitu.99@gmail.com}
\affiliation{Institute for Plasma Research, Bhat, Gandhinagar-382428, India}
\affiliation{Homi Bhabha National Institute, Anushaktinagar, Mumbai 400094, India}

\author{Hirakjyoti Sarma\,\orcidlink{0000-0003-0325-5650}}
\affiliation{Institute for Plasma Research, Bhat, Gandhinagar-382428, India}

\author{Ankit Dhaka\,\orcidlink{0000-0003-4425-9391}}
\affiliation{Atomic Semi Inc., Connecticut Street, San Francisco, CA 94107, USA}

\author{P. Bandyopadhyay\,\orcidlink{0000-0002-1857-8711}}
\affiliation{Institute for Plasma Research, Bhat, Gandhinagar-382428, India}
\affiliation{Homi Bhabha National Institute, Anushaktinagar, Mumbai 400094, India}

\author{A. Sen\,\orcidlink{0000-0001-9878-4330}}
\affiliation{Institute for Plasma Research, Bhat, Gandhinagar-382428, India}
\affiliation{Homi Bhabha National Institute, Anushaktinagar, Mumbai 400094, India}

\begin{abstract}
We present the first experimental evidence of inhomogeneous melting in a finite dusty plasma crystal confined in an anisotropic potential well. By systematically tuning the confinement anisotropy and applying controlled laser heating, distinct melting patterns are observed. Spectral-mode analysis based on Singular Value Decomposition of particle trajectories reveals that increasing laser power redistributes energy into specific collective modes, triggering localized structural destabilization. Molecular Dynamics simulations reproduce the observations and show that confinement-controlled mode coupling with laser heating governs the melting dynamics. These results establish geometric anisotropy as a key control parameter for inhomogeneous melting in finite coupled systems.
\end{abstract}

\maketitle
\textit{Introduction:} Finite systems composed of a limited number of interacting particles exhibit behavior fundamentally different from that of bulk matter due to strong boundary effects, discrete excitation spectra, and enhanced fluctuations \cite{Yan2016ExploringCrystals, Apolinario2008MultipleProperties, Home2011NormalPotentials}. These properties make finite ensembles valuable model systems for studying correlation-driven phenomena and collective dynamics at the mesoscopic scale. Finite interacting systems have been realized in a variety of experimental platforms including ultracold atomic gases \cite{IMMANUELBLOCH2005UltracoldLattices}, trapped-ion crystals \cite{DOnofrio2021RadialTrap}, electrons resting on liquid helium \cite{LeidererP.1981SurfaceHELIUM}, Wigner crystals \cite{Jean2002MacroscopicIslands}, and colloidal assemblies \cite{Tatarkova2002One-DimensionalParticles}. In such systems, the particles can self-organize in an ordered fashion under the joint influence of the interparticle interaction and the external trapping potential. Phase transitions leading to structural rearrangements \cite{Rancova2011NumericalSystems, Rancova2011StructuralClusters, Melzer2006ZigzagClusters,SchifferJ.P.1993Phase_transitions_in_ionic_crystals} including melting \cite{Boning2008MeltingSystems,Ferreira2006MeltingTrap,Tanaka2014MeltingCrystal, Ichiki2004MeltingPlasmas} have been widely studied for finite systems as they provide valuable insights into the microscopic dynamics of the transition process. 
\par
Symmetry plays an important role in the melting transition. A finite system with circular confining symmetry displays orientational (angular) melting - a phenomenon that has received considerable theoretical and experimental attention \cite{Filinov2001WignerSystems,Duca2023OrientationalParticles,Bubeck1999MeltingGeometry}. An asymmetry resulting from an anisotropic confining potential can lead to inhomogeneous melting that can proceed through non-uniform and geometry-dependent pathways. This was demonstrated in a theoretical study by Apolinario \textit{et al.} \cite{Apolinario2006InhomogeneousClusters} who performed detailed Molecular Dynamic simulations of classical charged particles trapped in an anisotropic parabolic potential. It was found that varying the eccentricity of the confining potential led to different melting patterns of the cluster. To the best of our knowledge, there has been no experimental verification of this interesting and fundamental finding. In this Letter, we present the first experimental demonstration of such inhomogeneous melting for a system of trapped charged dust particles embedded in a plasma.  Such a dusty plasma system provides a particularly advantageous platform to address this challenging problem because individual particle motion can be directly visualized while the confinement geometry can be varied with minimal perturbation to the ambient plasma conditions \cite{Sahu2025Confinement-drivenCrystal}.
By precisely tuning the confinement potential and using external laser heating, we demonstrate that the melting dynamics is governed primarily by confinement anisotropy and displays patterns that are akin to the theoretical ones. The evolution of the melting pathway is quantitatively resolved through mode-spectrum analysis based on Singular Value Decomposition and further supported by our own independent Molecular Dynamics simulations. Our results reveal the microscopic mechanism of such anisotropy-controlled melting in a finite coupled system and provide valuable insights that can prove useful for a wide range of confined mesoscopic platforms. 
\par
 \begin{figure}[htb]
    \centering
    \includegraphics[width=0.9\linewidth]{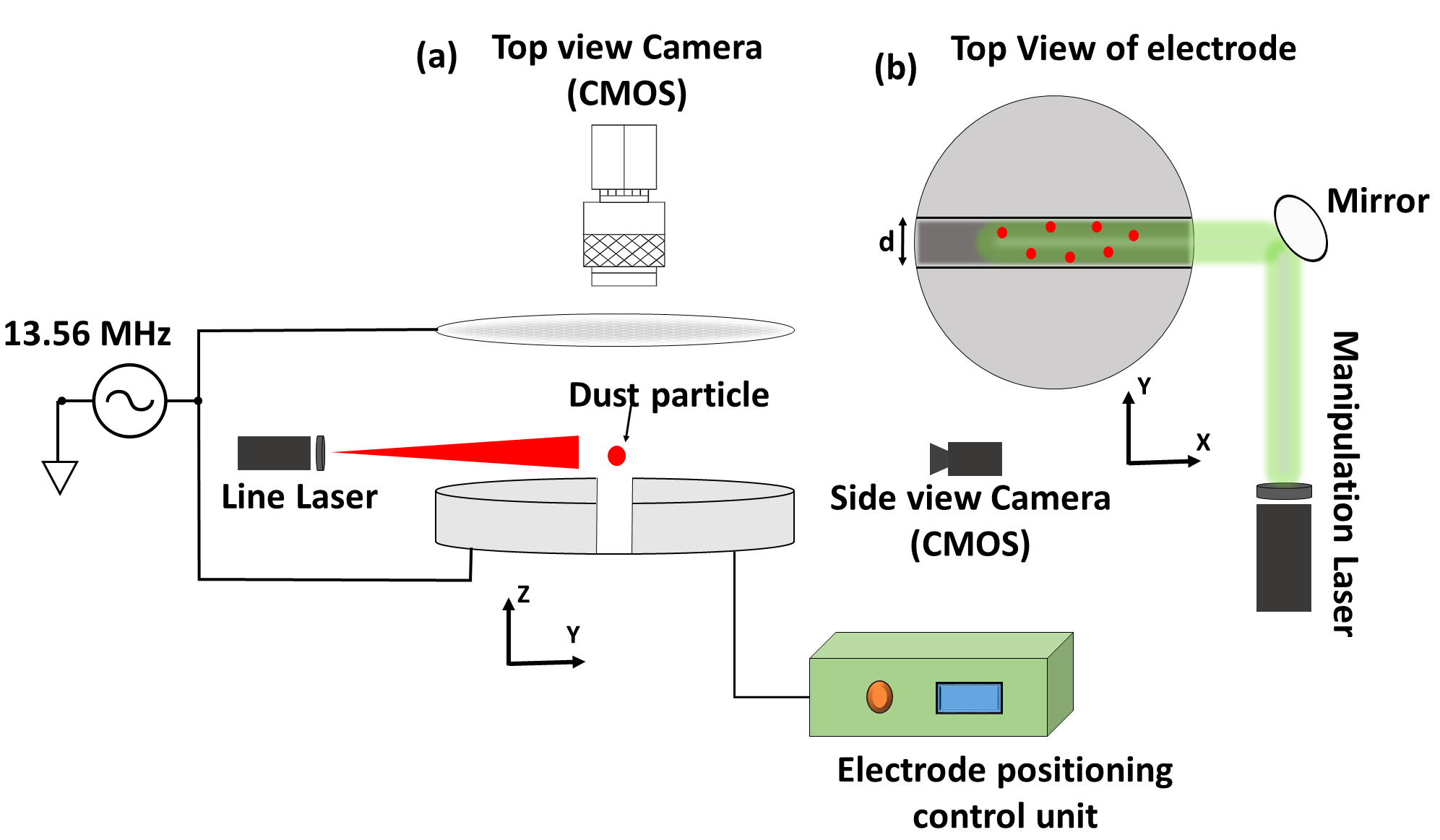}
    \caption{(a) Schematic illustration of the  experimental setup. (b) Top view of the lower electrode showing a central channel of width $d$, within which the particles are confined and externally driven by a collimated green laser beam. }
    \label{fig:schematic}
\end{figure}
\textit{Experiments:} The experiments are carried out in the Capacitively Coupled Dusty Plasma experimental (CCDPx) device \cite{Dhaka2025CapacitivelyPlasmas} as shown in Fig.~\ref{fig:schematic} . The plasma is generated between two parallel circular electrodes driven by a 13.56~MHz RF power source in push–pull configuration at an RF-power of $\sim$~1~W with V$_\mathrm{pp}=100$~V and pressure of 1.5~Pa. Plasma parameters such as electron temperature of $\sim 5$~eV, plasma density of $\sim 4\times10^{15}$~m$^{-3}$, and plasma potential of $\sim 9$~V are measured using an RF-compensated Langmuir probe \cite{Sudit1994RFDischargesb,Dhaka2025CapacitivelyPlasmas}. The confinement channel with variable width is created at the lower electrode using two D-shaped segments, one of which is actuated by a stepper motor controlled via a micro-controller. The separation between the electrodes can be tuned very precisely from 0 to $\sim$8~mm, enabling controlled variation of the in-plane confinement anisotropy without modifying bulk plasma parameters \cite{Dhaka2025CapacitivelyPlasmas, Sahu2025Confinement-drivenCrystal}. The anisotropy parameter, $\alpha$, defined as the ratio of the electric fields along and across the channel length, varies from $1 \sim~0.01$ with the increase in the channel width \cite{Sahu2025Confinement-drivenCrystal}. Seven mono-disperse melamine–formaldehyde (MF) particles of diameter $7.43 \pm 0.05~\mu$m are introduced into the plasma. These particles acquire a negative charge and levitate near the sheath-plasma boundary where the vertical electric force balances the gravity and form a two-dimensional monolayer cluster. The radial confinement comes from the radial component of the anisotropic electric field. 
 \begin{figure}[b]
    \centering
\includegraphics[width=0.99\linewidth]{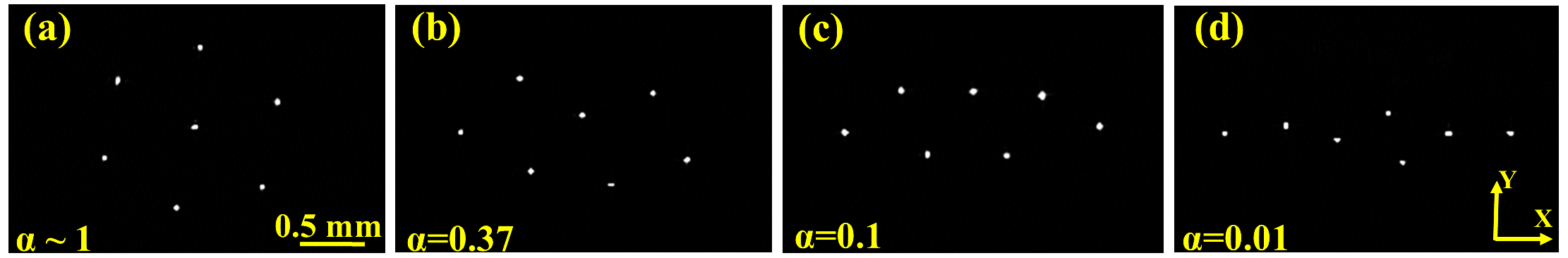}
    \caption{Experimentally obtained cluster of seven-particles for (a) $\alpha\sim1$,  (b) $\alpha = 0.37$, (c) $\alpha = 0.1$ and (d) $\alpha = 0.01$.}
    \label{fig:structures}
\end{figure}
The dust crystal is illuminated by a 630~nm, 100~mW line laser source equipped with a red filter for imaging.  Particle dynamics are recorded simultaneously using top and side-viewed high-speed CMOS cameras. Particle trajectories are obtained using the Trackpy package \cite{mTrackpy}, enabling quantitative characterization of the collective dynamics. A controlled melting is induced in the dust cluster using a collimated 532~nm laser beam (maximum power upto 1~W) with the help of a lens and mirror optical system that ensures spatially uniform external heating. \\
 \begin{figure}[b]
    \centering
\includegraphics[width=0.9\linewidth]{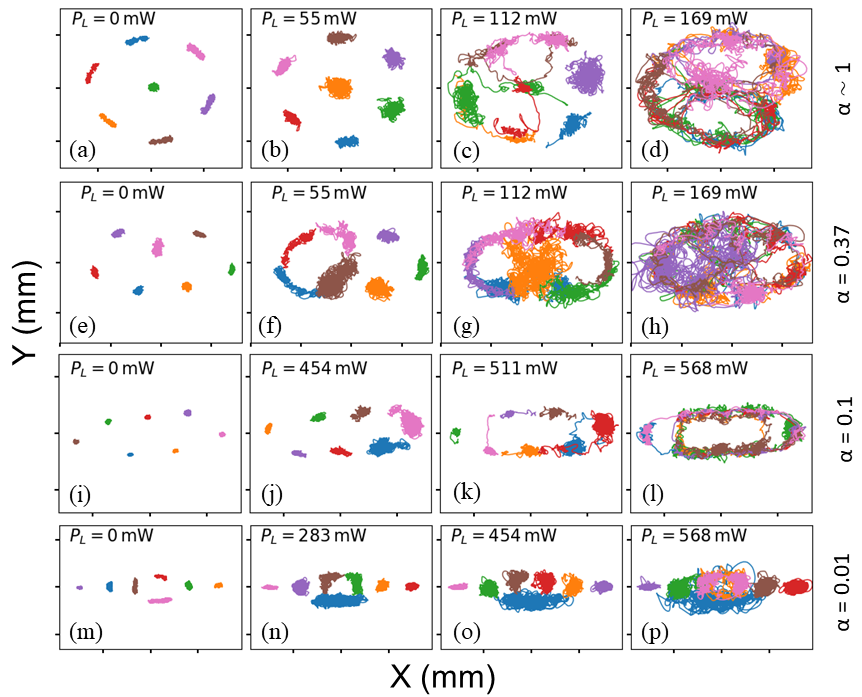}
    \caption{Particle trajectories of a nearly isotropic crystal containing seven particles ($\alpha \sim 1$) under progressively increasing laser power (a–-d); for $\alpha = 0.37$ (e–-h); for $\alpha = 0.1$ (i–-l); and for $\alpha = 0.01$ (m–-p). In each panel, trajectories are recorded over 56 s. 
 The distance between adjacent ticks is 1~mm.}
    \label{fig:trajectories}
\end{figure}
\textit{Results:} With the increase of channel width or decrease of anisotropy parameter ($\alpha$), the equilibrium configuration evolves from a two-dimensional circular cluster to an elliptical structure, then to a quasi-one-dimensional arrangement due to the underlying mechanism of sheath dynamics as shown in Fig.~\ref{fig:structures} and discussed in details in Ref.~\cite{Sahu2025Confinement-drivenCrystal}. To induce melting for a given value of $\alpha$, the laser power is imparted precisely on the cluster while keeping the discharge parameters constant. The experiments are repeated over a range of $\alpha$ values and laser powers.\par
 To investigate the melting scenario, the positions of the particles are overlapped for $\sim 56$~s for four different values of $\alpha$ and laser powers as shown in Fig.~\ref{fig:trajectories}. The calibrated laser power is imparted on the dust crystal from the right side. It is observed from the particle trajectories that the dust cluster undergoes a transition in a spatially inhomogeneous manner. For $\alpha \sim 1$, all the particles vibrate about their mean positions (see Fig~\ref{fig:trajectories}(a--b)) and later with increasing laser power, they exhibit a collective behavior that is manifested in discernible spatiotemporal patterns as illustrated in Fig.~\ref{fig:trajectories}(c--d).  At $\alpha = 0.37$, the particle dynamics becomes anisotropic and is accompanied by pattern formation for higher laser powers (see Fig.~\ref{fig:trajectories}(f--h)). The transition begins with a loop formation in the left region of the cluster (Fig.~\ref{fig:trajectories}(f)), followed by successive angular and radial excursions (Fig.~\ref{fig:trajectories}(g--h)) for $P_L > 112$~mW, where the particles organize into two-loop structures that are very similar to the earlier simulation study \cite{Apolinario2006InhomogeneousClusters}. 
 \begin{figure}[t]
    \centering
    \includegraphics[width=0.7\linewidth]{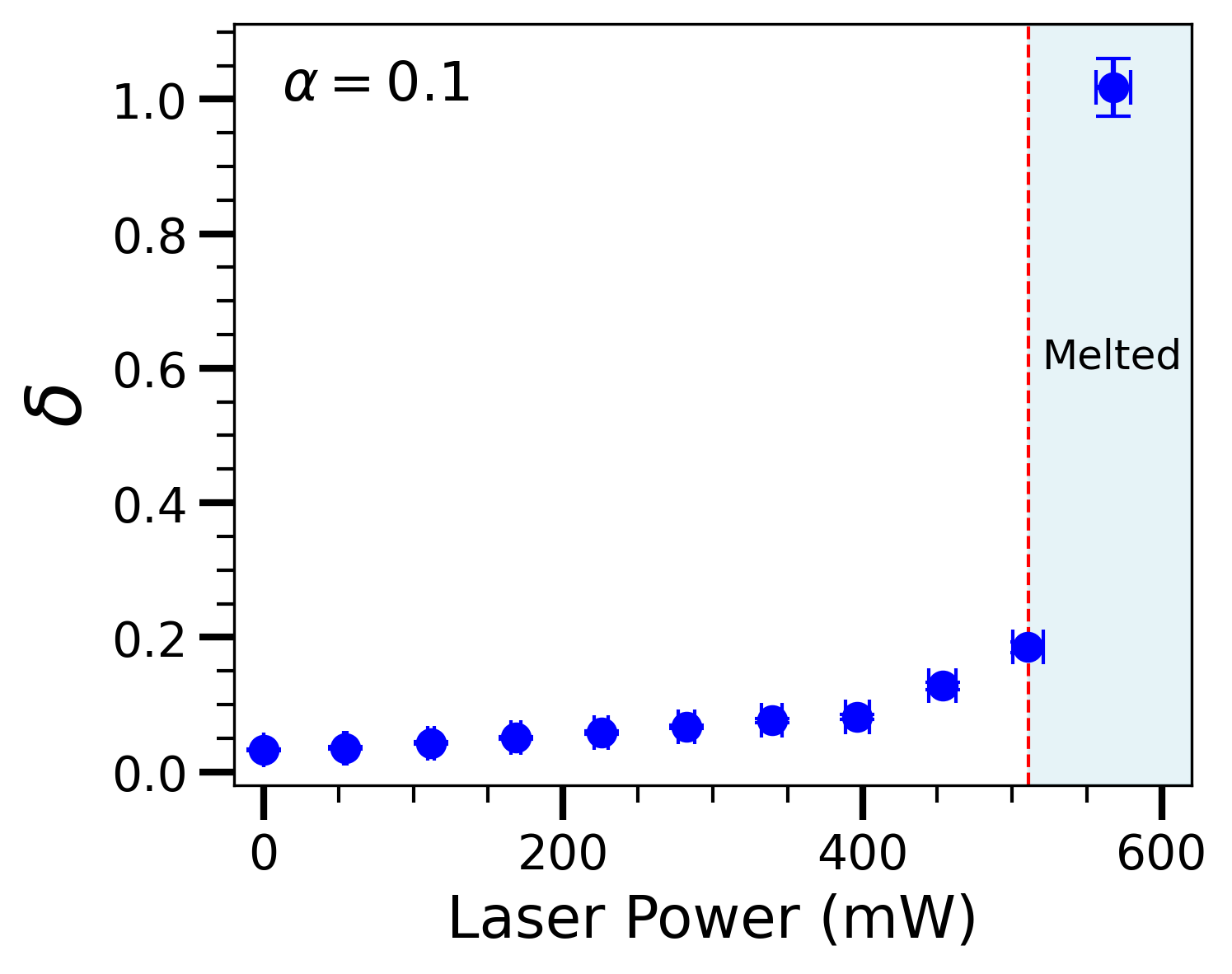}
    \caption{Variation of the Lindemann's parameter ($\delta$) with laser powers for $\alpha =0.1$. The vertical red dashed line is indicating the transition point at which the crystal undergoes melting.}
    \label{fig:u_x_fluctuation}
\end{figure}
Upon decreasing the anisotropy parameter further to $\alpha = 0.1$, the onset of pattern formation shifts to a higher laser power. At $P_L = 511\,\mathrm{mW}$, the cluster develops a localized pattern on the right side, while the leftmost particle continues to display thermal fluctuations and does not actively participate in the collective dynamics (Fig.~\ref{fig:trajectories}(k)). At $P_L = 568\,\mathrm{mW}$, the cluster exhibits a distinct multi-loop configuration, wherein the extreme end particles form two outer loops and the central particles organize into a compact inner loop (see Fig.~\ref{fig:trajectories}(l)). For a further reduction in the anisotropy to $\alpha = 0.01$, at an intermediate laser power of $P_L = 283\,\mathrm{mW}$, the extreme end particles vibrate about their mean position but the central region undergoes a pattern formation as illustrated in  Fig.~\ref{fig:trajectories}(n), which is akin to internal inter-shell melting as reported in the simulation study of Ref.~\cite{Apolinario2006InhomogeneousClusters}. At higher laser powers, the outermost particles increasingly participate in the dynamics, leading to a progressively global loss of structural order.\par
\begin{figure}
    \centering
    \includegraphics[width=\linewidth]{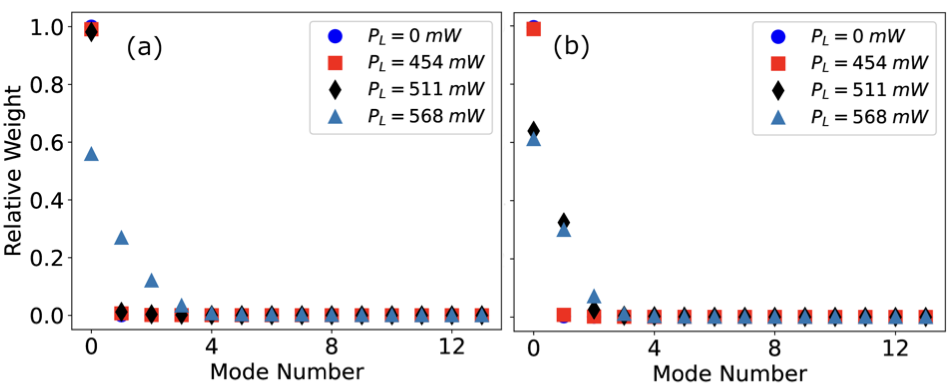}
    \caption{The relative modal weights at different values of the laser power obtained in (a) experiments and (b) simulation for $\alpha \sim 0.1$}

    \label{fig:sim_topo_relwt}
\end{figure}

To quantify the melting dynamics, we evaluate the Lindemann's parameter ($\delta$) defined as the ratio of the averaged particle displacement along the $x$-direction to the inter-particle distance ($a$) in the absence of laser power. Only the displacement along $x$ is considered because particle motion in the $y$-direction is strongly constrained by the extended sheath surrounding the channel. An increase in $\delta$ therefore indicates enhanced positional fluctuations, and melting is identified when these fluctuations reach a critical fraction of the mean inter-particle spacing (around $\sim$ 0.1), consistent with the Lindemann's criterion \cite{Khrapak2020LindemannDimensions}. Fig.~\ref{fig:u_x_fluctuation} shows the evolution of $\delta$ with increasing laser power for a fixed anisotropy parameter $\alpha$. A sharp increase in $\delta$ marks the loss of structural order and the onset of pattern formation, indicating the transition to the melted state. In the present experiment, this transition occurs at a critical laser power of approximately $511$~mW (see the third row of Fig.~\ref{fig:trajectories}). The Lindemann's criteria for other $\alpha$ values are discussed in detail in Sec.~I of the Supplemental Material \cite{supp}. The critical laser power required for melting is determined by the competition between the inter-particle interaction strength and the trapping potential. For larger values of $\alpha$, particle interactions dominate over the confinement potential, whereas decreasing $\alpha$ enhances the trapping strength along the $y$-direction. The resulting imbalance affects the stability of the crystalline configuration and governs the onset of melting. The observed inhomogeneous melting is therefore attributed to the combined influence of the laser radiation force and the anisotropic confinement potential. \\
\begin{figure}[t]
    \centering
    \includegraphics[width=1.0\linewidth]{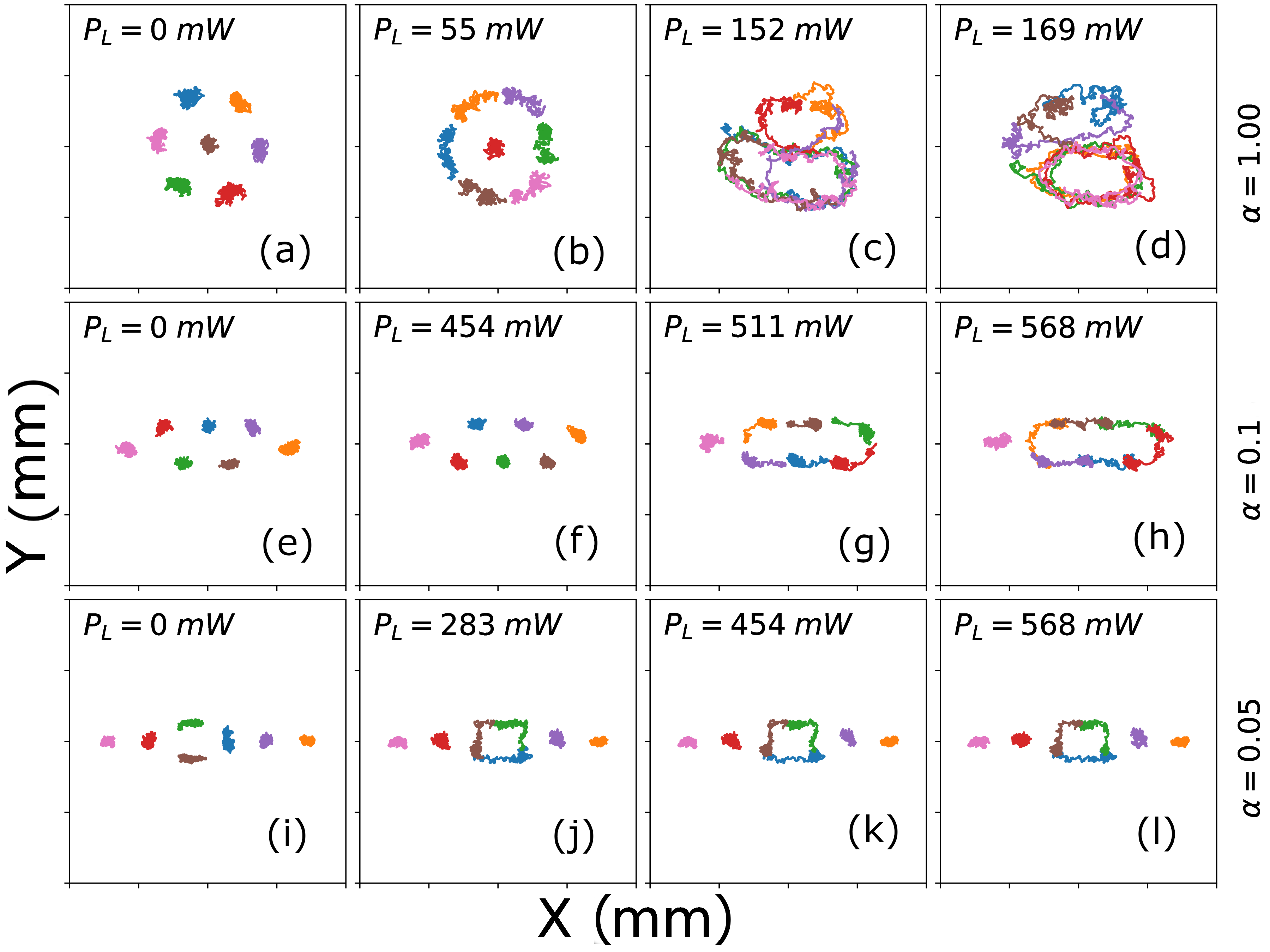}
    \caption{Particle trajectories from Langevin Dynamics simulations of a seven-particle anisotropically confined Yukawa system at three values of $\alpha$, corresponding to progressively increasing laser power: (a–d) $\alpha = 1$, (e-h) $\alpha = 0.1$, and (i–l) $\alpha = 0.05$. The trajectories are shown for $\sim$ ~56 seconds. The distance between adjacent ticks in the figures is 1~mm.}
    \label{fig:sim_traj}
\end{figure}
\textit{Langevin Dynamics simulation:} To complement our understanding of the melting mechanism, Langevin Dynamics simulations were performed on an anisotropically trapped system of charged point particles interacting through a repulsive Yukawa potential, using the Large-scale Atomic/Molecular Massively Parallel Simulator (LAMMPS) \cite{LAMMPS}. A detailed description of the simulations is given in Sec.~II of the Supplemental Material \cite{supp}. The trajectories obtained from the simulations at three different values of 
$\alpha$ are shown in Fig. \ref{fig:sim_traj}. For 
$\alpha = 1$ (Fig. \ref{fig:sim_traj}(a--d)), it is observed that at a relatively low but non-zero laser power ($P_L=55~mW$), a representative particle in the outer shell exhibits angular motion while the rest remain confined to the shell. At a higher laser power ($P_L =152$~mW), a particle in the outer-shell migrates towards the center of the trap, accompanied by a corresponding outward movement of the central particle. This behavior reproduces the experimental result of Fig. \ref{fig:trajectories}(a--d). Thus for the isotropic cluster angular melting precedes radial melting akin to earlier experimental findings \cite{Melzer2012InstantaneousClusters}. With a reduction in $\alpha$ to $0.1$, the melting threshold is shifted to a higher laser power in agreement with the experimental findings. The leftmost particle continues to oscillate about its mean position over the entire range of laser power considered, whereas the remaining particles form closed-loop trajectories, indicating a different dynamical response compared to the isotropic case. For $\alpha=0.05$ (Fig.~\ref{fig:sim_traj}(i--l)), it is observed that the central region of the cluster undergoes melting, while the particles located on the left and right sides continue to fluctuate about their respective mean positions. This behavior corresponds to what is known as internal inter-shell melting (Fig. \ref{fig:trajectories}(m--p)) \cite{Apolinario2006InhomogeneousClusters}. It is evident that the simulation trajectories closely resemble the experimental observations.\\
\begin{figure*}
\centering
    \includegraphics[width=0.9\textwidth]{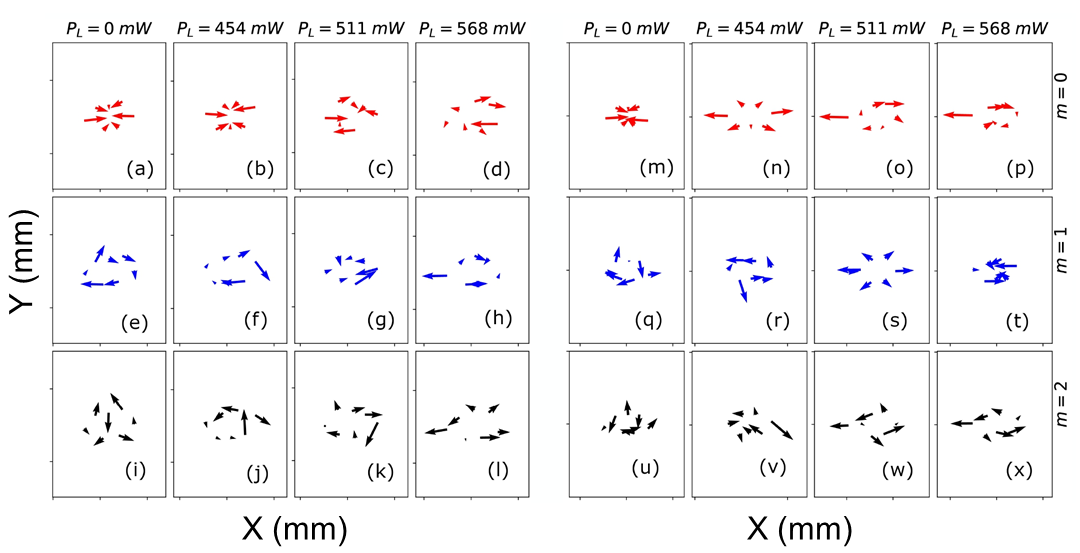}
    \caption{Spatial patterns ({\it topos}) of the first three modes for $\alpha \sim 0.1$ obtained from experimental particle trajectories (a--l)  and  simulation trajectories (m--x). The {\it topos} corresponding to $m=0,\;1\;\text{and}\;2$ are indicated by using red, blue and black arrows, respectively.}
    \label{fig:sim_expt_topo}
\end{figure*}
\textit{Mechanism of melting}: To elucidate the melting mechanism, the experimentally measured particle trajectories are decomposed into collective modes using the Singular Value Decomposition (SVD) method (for details, see Sec.~III of the Supplemental Material \cite{supp} and Ref.~\cite{dudok1994biorthogonal}). In this approach, the spatiotemporal dynamics are represented as separable spatial and temporal components, enabling identification of the dominant collective motion of the cluster. The first three SVD spatial patterns ({\it topos}) obtained from the experimental trajectories at four different laser powers for 
$\alpha=0.1$ are shown in Fig.~\ref{fig:sim_expt_topo}(a–l), together with the corresponding patterns extracted from Molecular Dynamics simulations in Fig.~\ref{fig:sim_expt_topo}(m–x). At zero laser power, the dominant modes correspond to well-defined collective motions, with the $m=0$ mode exhibiting a breathing-type (radial) oscillation and the $m=1$ mode displaying a circulating (azimuthal) motion. The spatial structure of the particle trajectories at a given laser power is therefore determined by the corresponding SVD {\it topos}. With increasing laser power, the mode structure evolves significantly. Beyond a threshold value of the laser power, a redistribution of the signal energy among several modes is observed both in the experiment and the simulation from the change in the relative modal weights as shown in Fig.~\ref{fig:sim_topo_relwt}(a) and Fig.~\ref{fig:sim_topo_relwt}(b), respectively. This redistribution arises from enhanced coupling between the SVD modes, causing the initially independent collective modes to interact. As a consequence, the spatial patterns gradually lose their pure form and evolve into mixed structures that combine features of multiple modes. Molecular Dynamics simulations reproduce this behavior and show a similar redistribution of modal energy with increasing laser power at fixed $\alpha$. These results indicate that laser-driven enhancement of mode coupling destabilizes the crystalline order, leading to the observed inhomogeneous melting. For stronger confinement anisotropy (smaller $\alpha$), the redistribution of modal energy becomes significantly weaker, consistent with the reduced tendency toward melting observed in this regime (see Sec.~III(b) of the Supplemental Material \cite{supp}).\\
\textit{Conclusion:} In summary, we present the first experimental observation of anisotropy-driven inhomogeneous melting in a finite dusty plasma cluster through external laser radiation heating. A seven-particle cluster is studied under a fixed set of discharge conditions, while the confinement anisotropy is varied through controlled adjustment of the channel width without modifying the background plasma condition. The cluster exhibits distinct melting pathways characterized by spatially inhomogeneous pattern formation. With increasing anisotropy, the system exhibits multiple melting pathways: purely radial, angular melting followed by radial melting and internal inter-shell melting. Complementary Langevin Dynamics simulations reproduce the experimental observations. Analysis based on Singular Value Decomposition reveals that the emergent structures originate from laser-driven redistribution of energy among collective modes, leading to enhanced mode coupling and destabilization of the crystalline order. These results identify confinement anisotropy as a key control parameter governing inhomogeneous melting dynamics in finite coupled systems and provide valuable fundamental insights that could be relevant to other confined mesoscopic systems.\\\\
\textit{Acknowledgments:} A.S. acknowledges the Indian National Science Academy for the INSA Honorary Scientist position. All the computations were performed on the HPC cluster ANTYA  at the Institute for Plasma Research, Gandhinagar, India.\\\\
\textit{Data availability:} The data that support the findings of this work are available from the corresponding author upon reasonable request.

\bibliography{references_monalisha}

\end{document}